\documentclass{PoS}

\usepackage{amsmath}
\usepackage{amssymb, fontenc, times, mathptmx, graphicx}
\usepackage[square,numbers]{natbib}

\title{Time dependent modeling of electron acceleration and cooling during blazar flares}

\ShortTitle{Modeling of electron acceleration and cooling during blazar flares}

\author{\speaker{A. Dmytriiev}, H. Sol, A. Zech\\
        Observatoire de Paris, CNRS, PSL Research University, LUTH, 5 Place J. Janssen, 92195 Meudon, France\\
        E-mail: \email{anton.dmytriiev@obspm.fr}}

\abstract{We present a new time-dependent leptonic code that we developed to model the varying multiwavelength (MWL) emission during blazar flares. In our modeling, we assume that the blazar emission originates from a plasma blob located in the jet, and that relativistic electrons are injected into the blob and may undergo stochastic (Fermi II) or shock (Fermi I) acceleration. We numerically solve the kinetic equation for electron evolution in the blob, taking into account particle injection, escape, acceleration and radiative cooling. In order to calculate the spectral energy distribution (SED) of the blob emission we assume a synchrotron self-Compton (SSC) scenario, including also synchrotron self absorption and gamma-gamma absorption processes. Our code computes the evolution of the electron spectrum and of the associated broad-band SED. As a first application, we attempt to connect the continuous, steady-state emission from the blazar Mrk 421 with a flare observed in February 2010, using a minimal number of free parameters in a two-zone scenario in which a turbulent region is present around the emitting zone. Mrk 421 is a high-synchrotron-peaked (HSP) BL Lac, and one of the brightest extragalactic $\gamma$-ray sources in the Very High Energy (VHE) $\gamma$-ray band. It is also the closest TeV emitting blazar to the Earth (redshift z=0.031).}

\FullConference{36th International Cosmic Ray Conference -ICRC2019-\\
		July 24th - August 1st, 2019\\
		Madison, WI, U.S.A.}

\begin{document}

\section{Introduction}
The non-thermal emission of blazars stretches across the entire electromagnetic spectrum from radio frequencies up to the VHE $\gamma$-ray range. The SED features two bumps. The first one spans from the radio band to the X-rays peaking in optical, UV or soft X-rays, and the second one extends up to TeV energies, showing a peak typically at $\sim$100 GeV. The lower energy component is explained as the synchrotron emission of relativistic electrons, while the high-energy bump in the SED is produced by inverse Compton (IC) scattering of electrons, interacting with a photon field. In a case when the seed photons are synchrotron photons emitted by the same electron population that up-scatters them, the scenario is referred to as synchrotron self-Compton \cite{maraschi}. This is the generally assumed scenario to describe BL Lac type blazars.  

Rapid and strong variability observed in all energy bands is a key property of blazars. The variability time scales depend on the frequency domain, and are ranging from minutes to months or even years. Of particular interest are flares in the VHE $\gamma$-ray band during which energy flux can increase by a factor of $\sim$10 within time scales ranging from $\sim$minutes to $\sim$days. A powerful tool to get an insight about the nature of blazar variability, is time dependent modeling of varying broad-band emission. We developed a new SSC code for time dependent modeling of variable MWL emission during flares of BL Lac objects, which allows to model the time evolution of the SED. We apply our code to an archival giant flare of Mrk 421 that occurred in February 2010.

\section{General time dependent model}

We assume a conventional leptonic scenario in which the radiation is produced in a spherical plasma blob of a radius $R_b$ embedded in a tangled homogeneous magnetic field $B$, and relativistically moving along the jet with a Doppler factor $\delta$. Electrons are continuously injected into the blob with some injection rate $Q_{inj}$ and may undergo stochastic or shock acceleration induced either by plasma turbulence in the blob or by a shock respectively. We use hard-sphere approximation ($q=2$) for particle-wave interactions in a turbulent medium, for which the characteristic time scale of stochastic acceleration process $t_{FII}$ is energy independent \cite{tramacere}. The population of the electrons radiates according to a one-zone synchrotron self-Compton scenario. The SSC emission can be affected by synchrotron self absorption and $\gamma$-$\gamma$ absorption processes. The electrons cool through radiative losses, and also escape the emission region with a characteristic time scale $t_{esc}$, which is also energy independent in the hard-sphere approximation case. Additionally, a turbulent region can be present around the blob, which can perturb its electron population and hence its emission.

The time evolution of the spectrum $N_e(\gamma,t)$ of electron population in the blob is described by a kinetic (Fokker-Planck) equation, which can be considered as a continuity equation in phase space. In our case it takes the following form \cite{tramacere}:

\vspace*{-3mm}

\begin{equation} \label{eq:1}
 		\dfrac{\partial N_e(\gamma,t)}{\partial t} = \dfrac{\partial}{\partial \gamma}\left( [\beta_{cool}\gamma^2 - 2D_0\gamma - a\gamma]N_e(\gamma,t)\right) + \dfrac{\partial}{\partial \gamma}\left(D_0\gamma^2 \dfrac{\partial N_e(\gamma,t)}{\partial \gamma}\right) - \dfrac{N_e(\gamma,t)}{t_{esc}} + Q_{inj}(\gamma,t)
 		\end{equation}

\hspace*{0.5mm} where {\small $\beta_{cool} = 4 \sigma_T (U_B + U_{rad}(\gamma,t)) / 3 m_e c$}, and the term $\beta_{cool}\gamma^2$ is the radiative cooling rate, $D(\gamma) = D_0 \gamma^2$ is the energy-diffusion coefficient, $D_0 = 1 / t_{FII}$ is the inverse of the characteristic timescale of the stochastic acceleration process \cite{parkpet1995}, $a = 1 / t_{FI}$ is inverse of characteristic timescale of shock acceleration, and $Q_{inj}(\gamma,t)$ is the injection function. Currently, we are only treating BL Lac objects and consider that the synchrotron cooling is dominant.

Based on this model, we developed a code for modeling multi-frequency emission during blazar flares. We numerically solve the kinetic equation by using a fully implicit difference scheme proposed by Chang and Cooper (1970) \cite{changcooper}, taking into account the processes mentioned above. Having electron spectrum evolution on a time grid, we compute at each time step the broad-band SED of the emission from the blob, associated with the electron spectrum. We implemented the expressions for intensities of synchrotron and SSC emission from \cite{cg1999}. We take into account the Extragalactic Background Light (EBL) absorption using a publicly available code \footnote{https://github.com/me-manu/ebltable}. In our modeling we use the EBL model by Dominguez \cite{dominguez}. The light curve is calculated via integration of the time dependent emission in a certain energy range.

\section{Application to the blazar Mrk 421: quiescent and flaring states}

\subsection{February 2010 flare of Mrk 421: the data}

Mrk 421 is highly variable in X-rays and $\gamma$-rays, especially in the VHE $\gamma$-ray range. During the period February 10-23, 2010 the source showed strong flaring activity in X-rays and $\gamma$-rays, with the flux reaching a peak on February 17, 2010. In the VHE $\gamma$-ray band, the flare was observed fully (rise and decay) with the High Altitude GAmma Ray (HAGAR) telescope array at energies above 250 GeV during February 13 - 19, and the TeV Atmospheric Cherenkov Telescope with Imaging Camera (TACTIC) during February 10 - 23, and partially (at the peak and/or after) by High Energy Stereoscopic System (H.E.S.S.) from 17th to 20th of February \cite{hessf}, and VERITAS during the peak of the flare on February 17 \cite{veritasf}. Various other instruments across the world monitored the outburst in different energy bands. In order to understand the nature of this flaring activity, we investigate the temporal characteristics of the variable emission of the source in optical, X-ray and $\gamma$-ray energy bands. We use near simultaneous archival dataset from \cite{shukla2012}, which comprises light curves measured with Swift-XRT and RXTE-PCA in X-rays, Fermi-LAT in $\gamma$-rays and HAGAR in the VHE $\gamma$-ray range, as well as the reduced optical V-band observations by SPOL at Steward Observatory. We complement the dataset with the TACTIC measurements in the VHE $\gamma$-rays which we take from \cite{singh}. The overall dataset spans the period February 10-26, 2010 (MJD 55237-55253).

\subsection{Modeling of the quiescent state of the source}

\begin{figure}[h]
\centering
\includegraphics[width=94mm]{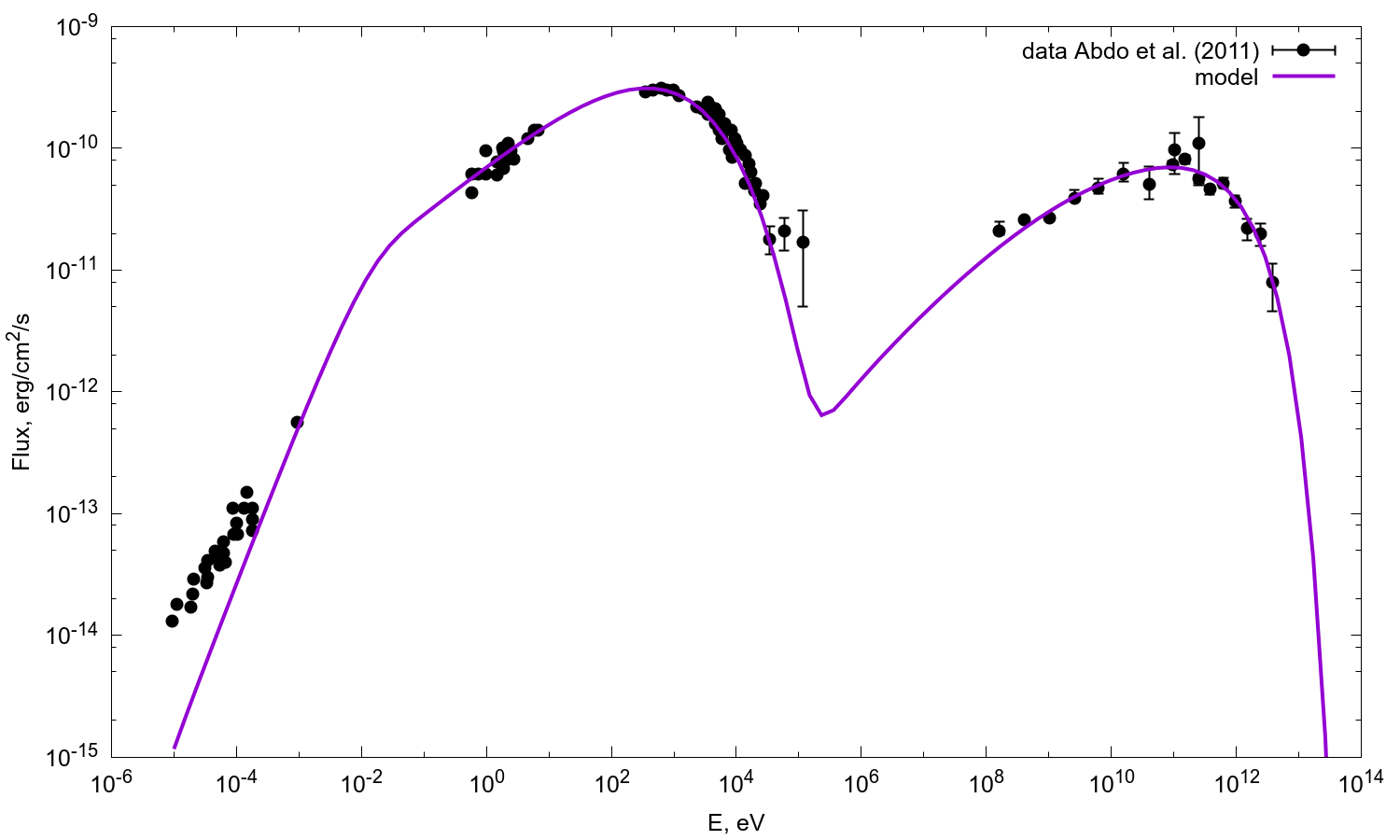}
\caption{Modeling of the broad-band quiescent emission from Mrk 421. Violet curve -- our model (asymptotically established SED, EBL not included), black points -- dataset from \cite{abdo2011} (EBL de-absorbed)}
\label{fig:qsm}
\end{figure}

In this work, flaring behavior is considered as a perturbation above the quiescent (steady) state of the source, which can have very different causes: injection rate change, a shock passing through the blob or vice versa, or suddenly enhanced turbulence, etc. To simulate flares, we need to firstly reproduce the non-flaring state of Mrk 421, and then introduce disturbances into the system. For our modeling of the steady state we use a dataset from \cite{abdo2011}, and suppose that pre-accelerated electrons coming presumably from the base of the jet (in a form of a steady flow) are continuously injected into the emission zone. The spectrum of these electrons is assumed to be a power law with exponential cutoff, generated by Fermi I acceleration. A time scale of particle escape $t_{esc} = 1 \, R/c$ is used. The electron population in the blob radiates according to the SSC scenario, and cools. The quiescent (steady) state of Mrk 421 in this scenario is achieved by asymptotically established balance between the gain processes (injection) and losses (cooling and escape). We fit the data points by varying physical parameters of the source. Fig. \ref{fig:qsm} shows the comparison between the asymptotic steady-state SED and the MWL measurements from \cite{abdo2011}, and the parameters are listed in Table \ref{table:1}. They are in a good agreement with \cite{abdo2011}. Also, the electron spectrum in the blob has similar slope (2.18 vs 2.2) and the cooling break position is comparable ($\gamma_{cool} \simeq 3 \cdot 10^5$). A flare can be launched by perturbing the system during a certain period of time as described hereafter.

\begin{table}[h]
\footnotesize
\centering
\begin{tabular}{ |c|c|c|c| } 
 \hline
 Parameter & Symbol & Our model & Abdo et al. (2011) \cite{abdo2011} \\ 
 \hline
  Magnetic Field [G] & $B$ & $0.038$ & $0.038$  \\ 
 \hline
  Comoving blob radius [cm] & $R_b$ & $2.46 \cdot 10^{16}$ & $5.2 \cdot 10^{16}$ \\
 \hline
 Doppler Factor & $\delta_b$ & $32$ & $21$  \\ 
 \hline
 Minimum Electron Lorentz Factor & $\gamma_{min}$ & $800$ & $800$  \\ 
 \hline
 Spectrum of injected electrons & $Q_{inj,q}$ & $ A \cdot \gamma^{-p_{inj}} \cdot exp(-\gamma/\gamma_{cut})$ & not defined  \\
 Injection spectrum normalization [$cm^{-3} s^{-1}$] & $A$ & $2.15 \cdot 10^{-3}$ & \\
 Injection spectrum slope & $p_{inj}$ & $2.18$ & \\
 Lorentz Factor of exp. cutoff in inj. spectrum & $\gamma_{cut}$ & $2.02 \cdot 10^5$ & \\
 \hline
\end{tabular}
\caption{Physical parameters of Mrk 421 steady state. Left -- our model, right -- stationary model by \cite{abdo2011}}
\label{table:1}
\end{table}

\subsection{Modeling of Mrk 421 February 2010 flare}

\subsubsection{One-zone scenario} \label{sssec:1zone}

To keep a minimal number of free parameters, we first consider that the flaring component emanates from exactly the same region as the continuous emission. We assume that the flare is caused by the interaction between a shock and the emitting blob and/or by turbulence affecting the entire emission region. However this scenario cannot adequately explain the data (see \ref{subsec:resonezone}). 

\subsubsection{Two-zone scenario} \label{sssec:2zone}

\begin{figure}[h]
\centering
\includegraphics[width=60mm]{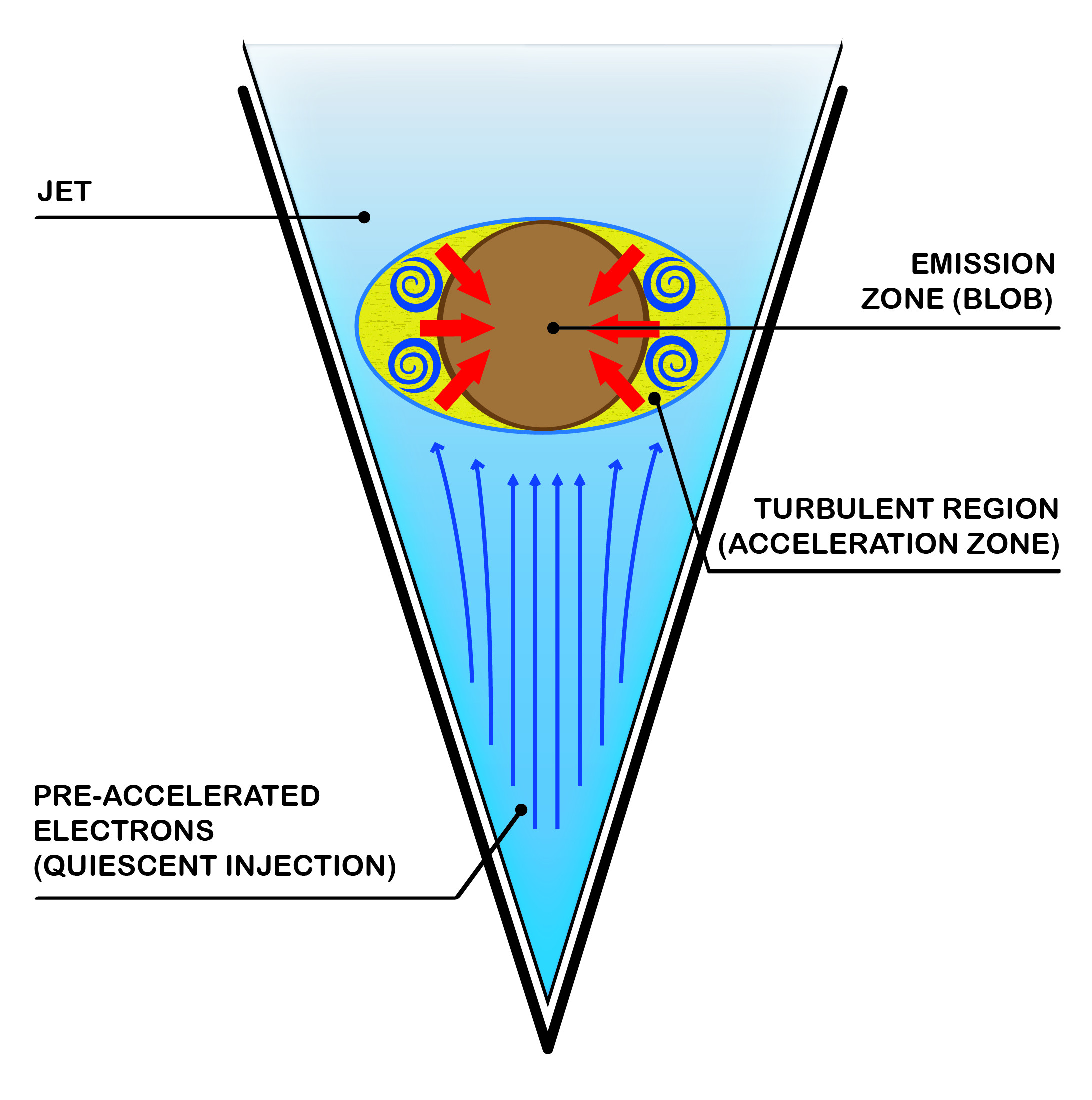}
\caption{Sketch  representing  the two-zone  scenario with a turbulent region around the blob. Red arrows show the flow of electrons from the acceleration zone to the blob.}
\label{fig:sketch}
\end{figure}

We propose a scenario where the observed increase of energy flux from the source is induced by inflow of electrons in the blob from the outside, additionally to the stream of pre-accelerated particles coming from the base of the jet (responsible for production of the quiescent emission). The light curves show that the source is strongly variable in X-rays, but weakly variable in the optical V-band. Such MWL behavior can result from injection of electrons with a harder spectrum, than the one of the electron population in the blob in the quiescent state ($N_{e,q} \propto \gamma^{-2.2}$). We suppose that this additionally injected population of electrons originates from an external non-radiative zone, and is produced via stochastic acceleration. We assume that shortly before the start of the flare, a turbulent zone forms around the blob due to various perturbations near the boundary layer which cause laminar-turbulent transition in the overflowing plasma. After the turbulence around the blob has been developed, pre-accelerated electrons arriving from the base of the jet are entering the turbulent zone and undergo stochastic acceleration. The particles escape the turbulent region, and a fraction of them is injected inside the blob (see Fig. \ref{fig:sketch}). The magnetic field in the acceleration zone is supposed to be small, rendering any emission from it negligible. 

The additional electrons sum up to the electron population in the blob, causing flux increase. Since these particles have a harder spectrum than the quiescent population, an excess of electrons is created at high Lorentz factors, leading to a flare with stronger variability in hard X-rays compared to the optical band.

\section{Modeling results}

\subsection{One-zone scenario} \label{subsec:resonezone}

We try to reproduce the observational data with the model from \ref{sssec:1zone}. For that, we supplement our modeling of the steady state with a perturbation: shock and/or stochastic acceleration of electrons in the emitting blob, and vary the parameters of the shock and turbulence ($t_{FI}$ and $t_{FII}$ respectively), keeping the physical parameters of the emitting blob constant. It appeared to be impossible to fit the data with any values of the parameters, since both acceleration processes perturb significantly the SED in optical V-band if we achieve the observed flux enhancement in the X-rays, while little variability is seen in the optical data.     

\subsection{Two-zone scenario} \label{subsec:restwozone}

We try to overcome the issue encountered in the one-zone scenario by introducing an additional acceleration zone. We use the light curves to extract the information about the additional supply of particles to the emitting blob. Using an approximation, that an electron with the Lorentz factor $\gamma_0$ emits only at the synchrotron frequency $\nu_s \simeq 1.2 \times 10^6 \, B \gamma_0^2 \delta \, / \, (1+z)$, the intensity of synchrotron emission at the frequency $\nu_s$ is proportional to the number density of particles at the Lorentz factor $\gamma_0$: $I_{s}(\nu_s) \propto N_e(\gamma_0)$. The light curve in the optical V-band, and four light curves in the X-rays display variations of the synchrotron flux, and thus reflect the evolution of electron population in the blob at the five Lorentz factors corresponding to the energy ranges. We evaluate the quiescent fluxes in the five bands using the steady-state SED, fit the reduced light curves (measured flux minus the quiescent value) with a simple exponential rise and decay, and retrieve the time evolution of the additional electron population at the five Lorentz factors from the light curves of the flaring emission. In order to determine the varying additional electron injection that causes the recovered behavior of the additional electron population, we use the kinetic equation (Eq. \ref{eq:1}) that links them. The slope of the injection spectrum was found to harden during the rise of the flare, and remain constant during its decay. This behaviour can be attributed to an acceleration process.

We model the two zones with our code. First we model electron acceleration in the turbulent region. We use the same injection spectrum as for production of the steady state (see Tab. \ref{table:1}), however with a slightly different norma\-lization, since the turbulent zone is much smaller. We rescaled the normalization requiring that the particle flux (number/cm$^2$/s) in the flow near the blob and near the turbulent zone are the same, as it is a single continuous flow. Another important constraint is that the size of the acceleration zone and its magnetic field both should be smaller than in the emitting blob, since the radiation from the turbulent region is assumed to be subdominant. We roughly estimate the physical parameters of the acceleration zone by approximately reproducing the reconstructed time evolution of the additional injection spectrum. We suppose that the turbulence in the acceleration region abruptly starts, lasts for a certain time $t_{l,az}$ and then immediately ends. During the life time of the turbulence the magnetic field is tangled and the particles can be entrapped inside the region, and the characteristic of the turbulence, the time scale of the stochastic acceleration $t_{FII,az}$ is constant. Then we model the emission from the blob. We set the electron spectrum in quiescent state as initial condition, and use the same physical parameters as for the steady state of the source. We are searching for the parameters of the turbulent region such that the accelerated electrons that escape into the emitting zone can explain the flaring emission. We fine-tune the parameters in a way that the simulated light curves have to match the data.

\begin{table}[h]
\footnotesize
\centering
\begin{tabular}{ |c|c|c| } 
 \hline
 Parameter & Symbol & Value  \\ 
 \hline
  Magnetic Field [G] & $B_{az}$ & $0.027$   \\ 
 \hline
 Comoving effective size [cm] & $R_{az}$ & $5.51 \cdot 10^{15}$ \\
 \hline
 Time scale of the stochastic acceleration [d] & $t_{FII,az}$ & $43$ $R_{az}/c$ $\approx 91.5$ d. (acc. zone frame) \\
 \hline
 Time scale of the particle escape [d] & $t_{esc,az}$ & $18$ $R_{az}/c$ $\approx 38.3$ d. (acc. zone frame) \\
 \hline
 Life time [d] & $t_{l,az}$ & $4.65$ d. (observ. frame)\\
 \hline
 Fraction of particles injected in the blob [] & $f_{inj}$ & $3$ \%\\
 \hline
\end{tabular}
\caption{Physical parameters of the turbulent region}
\label{tab:turbreg}
\end{table}

\begin{figure}[h]
\centering
\includegraphics[height=33mm]{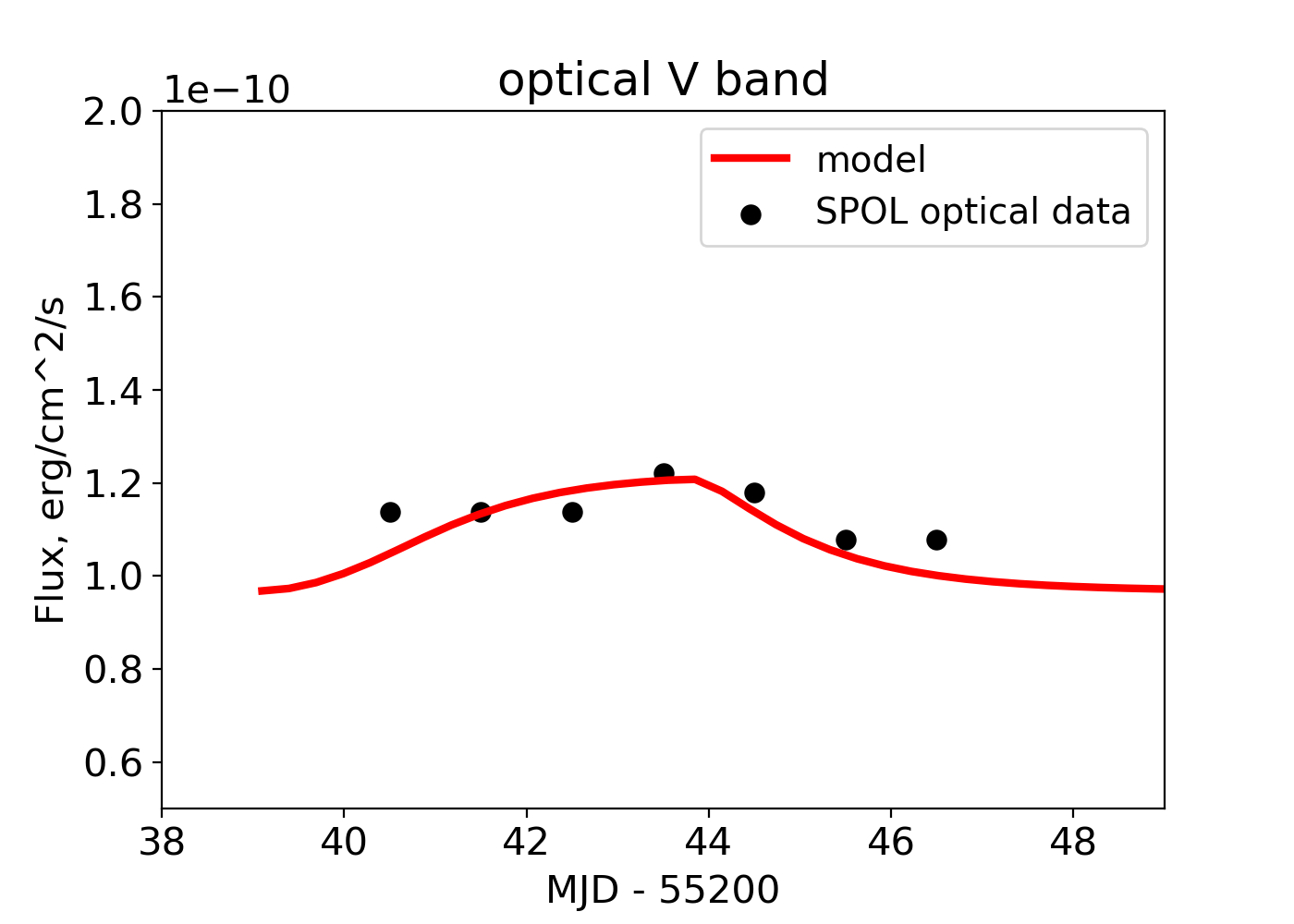} \hspace*{0mm}  \includegraphics[height=33mm]{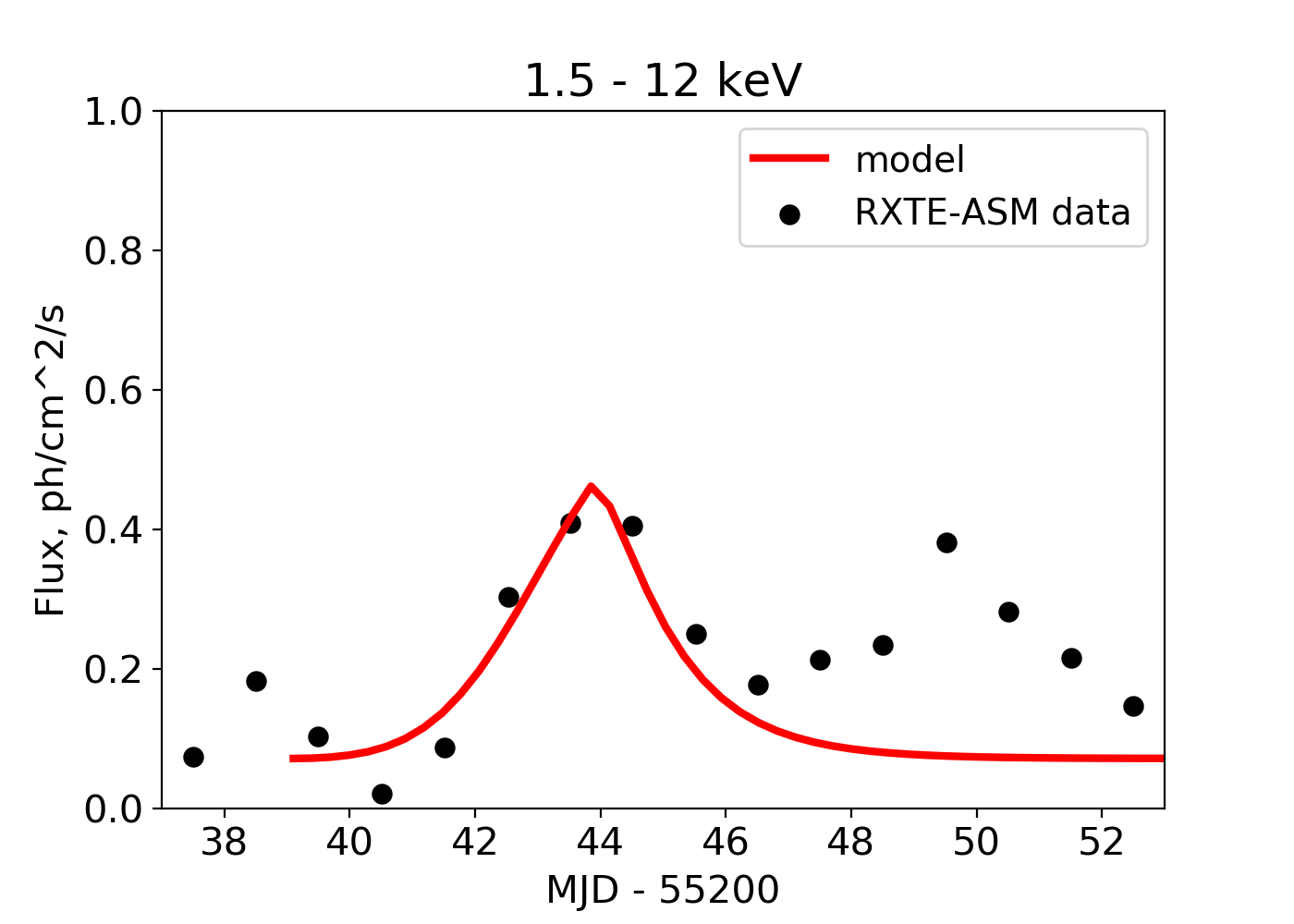}

\vspace*{0mm}

\includegraphics[height=33mm]{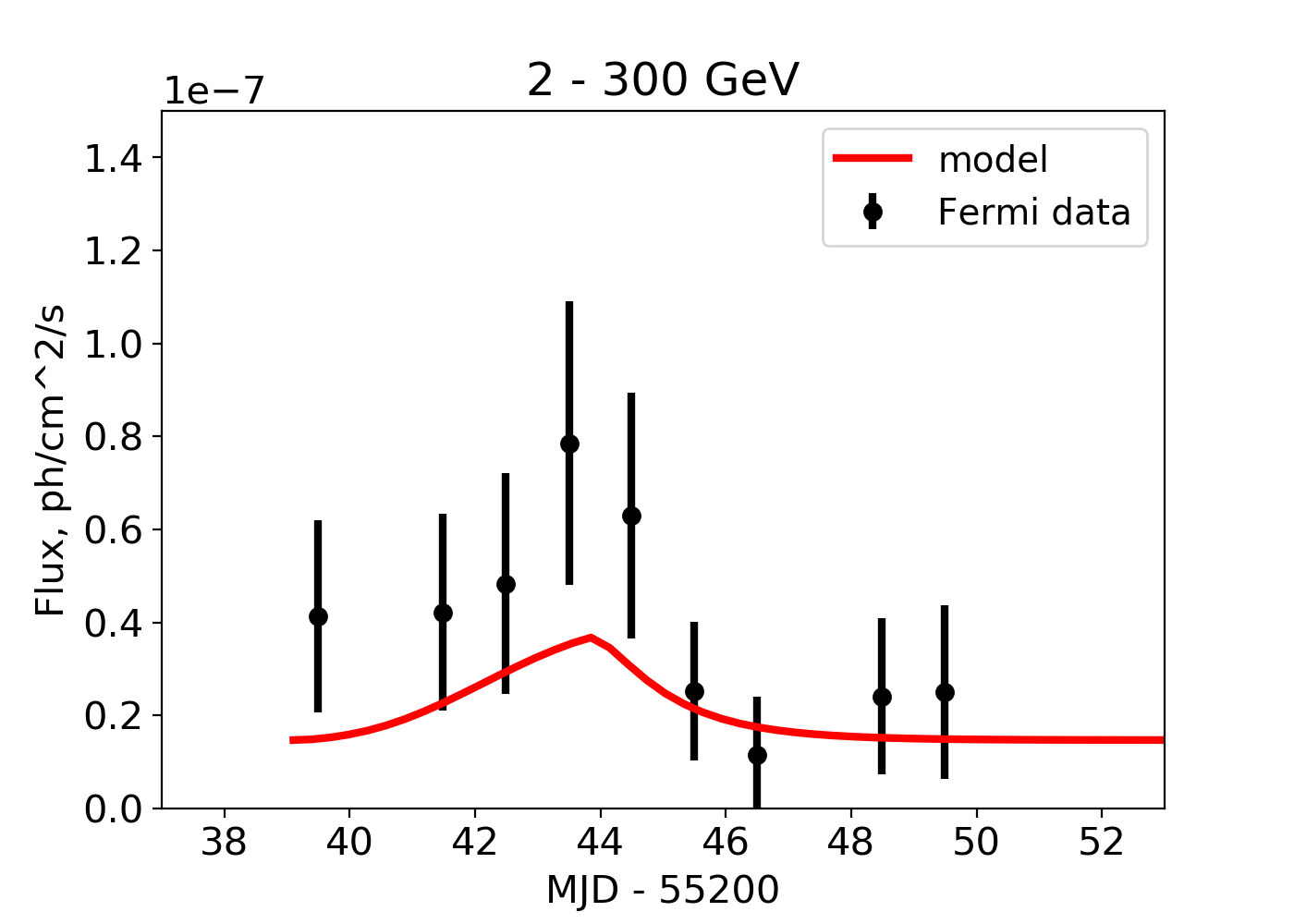} \hspace*{0mm} \includegraphics[height=33mm]{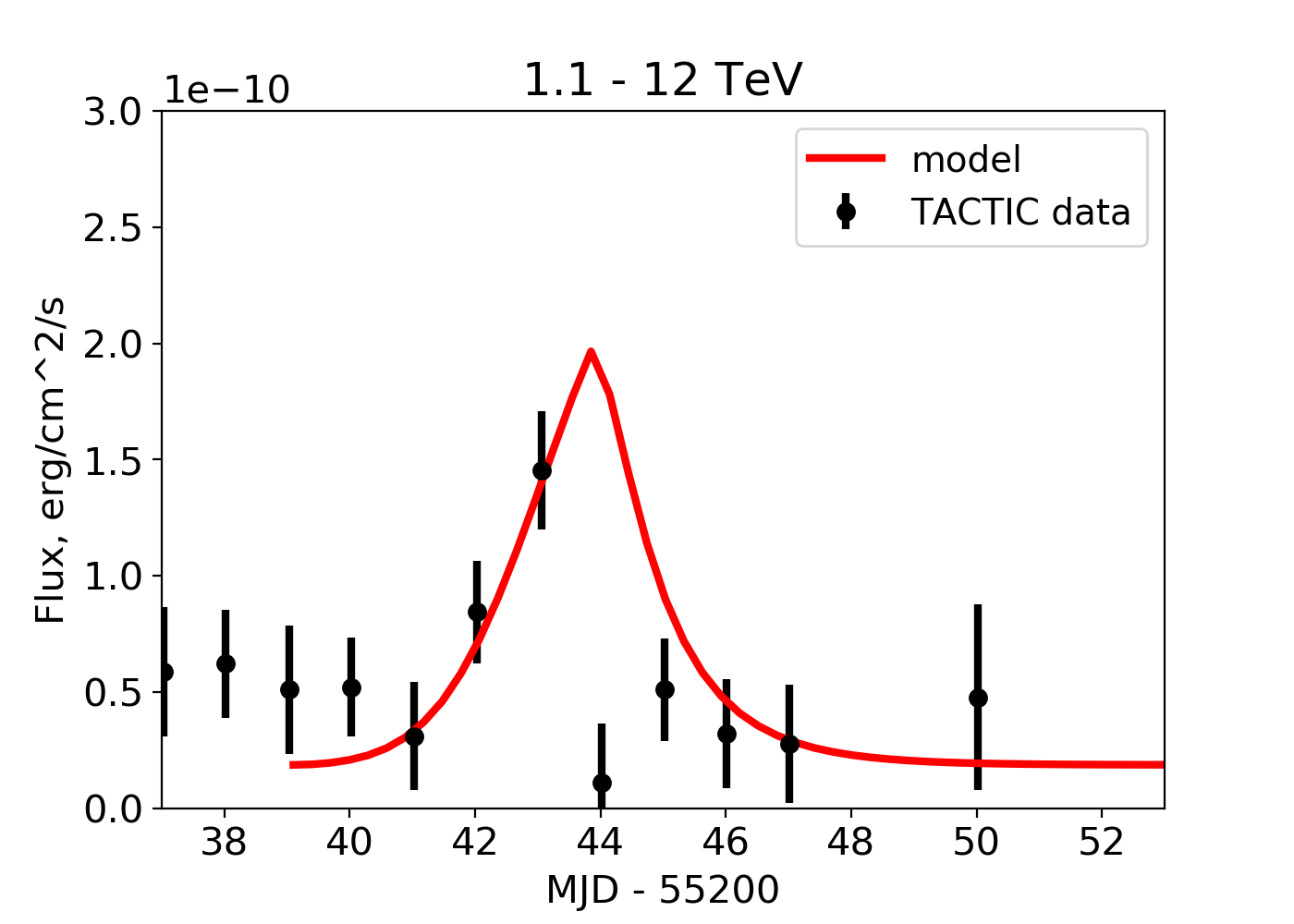}

\caption{Comparison of model light curves and archival MWL data taken from \cite{shukla2012} and \cite{singh}}
\label{fig:lcs}
\end{figure}

\begin{figure}[h]
\centering
\includegraphics[height=64mm]{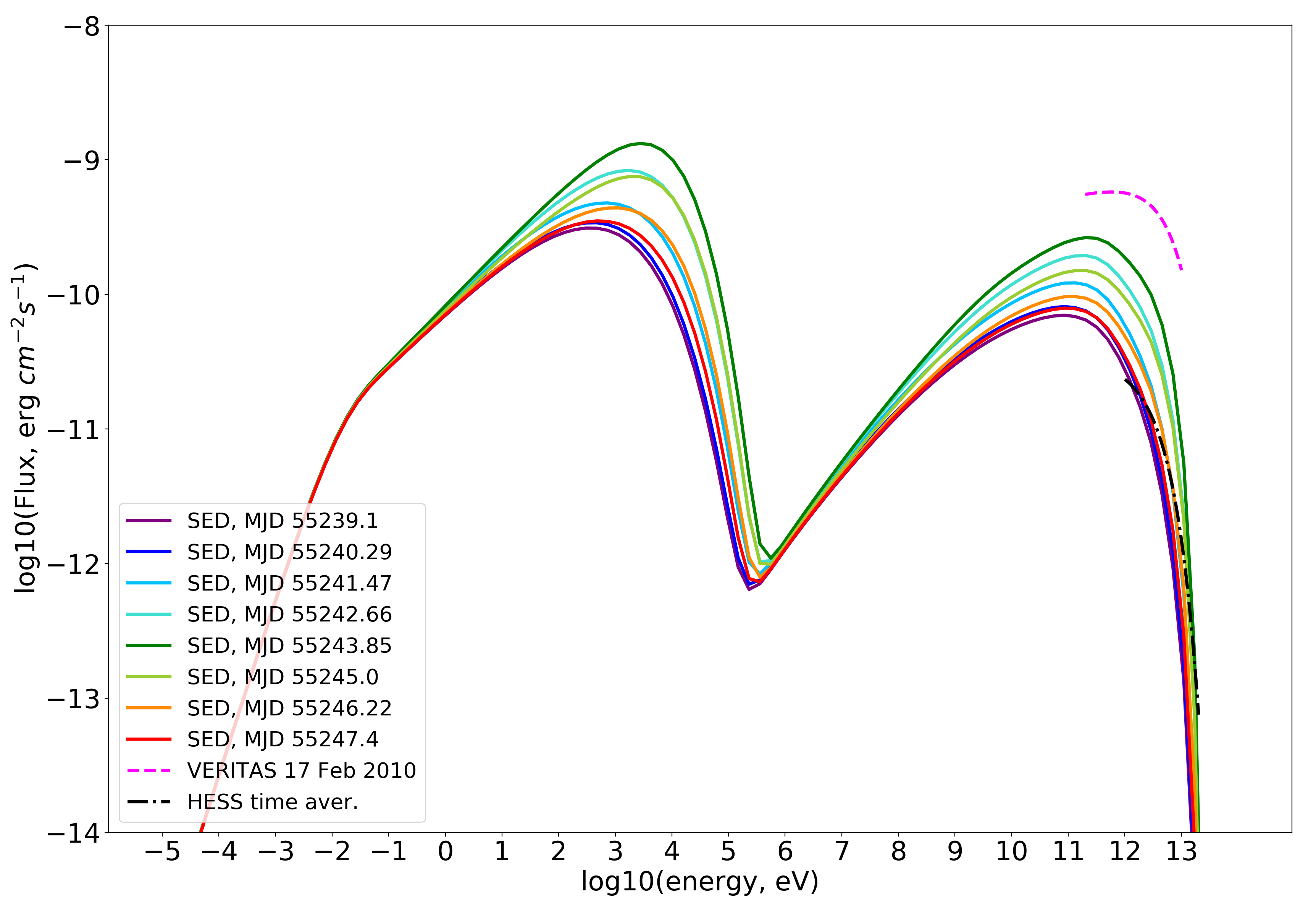}
\caption{Solid lines: Modeled time evolution of the SED (advancing from violet to red) during the flare. Magenta dashed line: peak measurement of the SED by VERITAS from \cite{veritasf}. Black dash-dotted line: H.E.S.S. time averaged SED during the flare decay from \cite{hessf}}
\label{fig:sedevol}
\end{figure}

Fig. \ref{fig:lcs} shows the comparison of the model representations of the light curves and a subset of the available data, and the parameters are presented in the Table \ref{tab:turbreg}. The evolution of the SED, with the VERITAS and H.E.S.S. data imposed are shown in Fig. \ref{fig:sedevol}. We arrive at an overall good representation of the MWL light curves and the SED, but the present model still underpredicts the $\gamma$-ray emission during the flaring activity. In GeV energy band, the simulated light curves have a correct shape, but lower flux than in the data by a factor of $\sim$ 3. Same can be also seen in $\sim$100 GeV range by comparison of the peak SED measurement by VERITAS and the model. In the TeV energy range the description of the flux evolution is rather satisfactory. To verify our assumption about low contribution of the turbulent region emission, we evaluate the ratio of the luminosity of the acceleration zone to the luminosity of the emitting zone. Luminosity $L$ [$erg/s$] $\propto A_{inj} R^3 B^2$, where $A_{inj}$ is the injection spectrum normalization [$cm^{-3} s^{-1}$]. The ratio of luminosities is then $\sim B_{az}^2 R_{az}^3 / B^2 R_b^3$ $\approx 10^{-3}$, thus the emission from the turbulent zone is indeed negligible.

\section{Conclusions}
We attempt to explain the flaring activity of Mrk 421 observed in February 2010 with one-zone and two-zone leptonic scenarios. We find that within a one-zone scenario it is not possible to achieve a flaring state of the source starting from a quiescent state by a perturbation (turbulence and/or shock) localized to the emitting blob. A two-zone scenario with a single emission zone and an additional acceleration zone that injects a hard particle spectrum seems to lead to overall satisfactory results. However, the representation of the $\gamma$-ray data is still imperfect, and requires further investigation. It might for instance indicate that the flaring state emission originates not exactly from the same zone as the steady state emission. 

\section{Acknowledgements}

We express gratitude to Cosimo Nigro (DESY, Zeuthen, Germany) for letting us use and extend his code that solves the kinetic equation with the Chang and Cooper numerical scheme. We also thank Manuel Meyer (Stanford University) for development of the very useful publicly available code allowing to compute attenuation factors due to the EBL absorption.

\end{document}